\documentclass[runningheads]{llncs}
\usepackage{graphicx}
\usepackage{url}
\usepackage{xcolor}
\usepackage{tabu} 
\usepackage{booktabs}   
\usepackage[caption=false]{subfig}

\begin{document}
\title{Interface Design for HCI Classroom: From Learners' Perspective}

\author{Huyen N. Nguyen\orcidID{0000-0001-6554-2327} \and
Vinh T. Nguyen\orcidID{0000-0002-1300-3943} \and
Tommy Dang\orcidID{0000-0001-8322-0014}}
\authorrunning{H. N. Nguyen et al.}
\institute{
Texas Tech University, Lubbock TX 79409, USA\\
\email{\{huyen.nguyen,vinh.nguyen,tommy.dang\}@ttu.edu}}
\maketitle              % typeset the header of the contribution
\begin{abstract}

Having a good Human-Computer Interaction (HCI) design is challenging. Previous works have contributed significantly to fostering HCI, including design principle with report study from the instructor view. The questions of how and to what extent students perceive the design principles are still left open. To answer this question, this paper conducts a study of HCI adoption in the classroom. The studio-based learning method was adapted to teach 83 graduate and undergraduate students in 16 weeks long with four activities. A standalone presentation tool for instant online peer feedback during the presentation session was developed to help students justify and critique other's work. Our tool provides a sandbox, which supports multiple application types, including Web-applications, Object Detection, Web-based Virtual Reality (VR), and Augmented Reality (AR). After presenting one assignment and two projects, our results showed that students acquired a better understanding of the Golden Rules principle over time, which was demonstrated by the development of visual interface design. The Wordcloud reveals the primary focus was on the user interface and shed some light on students' interest in user experience. The inter-rater score indicates the agreement among students that they have the same level of understanding of the principles. The results show a high level of guideline compliance with HCI principles, in which we witnessed variations in visual cognitive styles. Regardless of diversity in visual preference, the students presented high consistency and a similar perspective on adopting HCI design principles. The results also elicited suggestions into the development of the HCI curriculum in the future.

\end{abstract}

\section{Introduction}
\label{sec:intro}

HCI has a long, rich history, and its origin can be dated back to the 1980s with the advent of personal computing. As such, computers were no longer considered an expensive tool, and room-sized dedicated to experts in a given domain. Consequently, the need to have an easy and efficient interaction for general and untrained users became increasingly vital for technology adoption. In recent years, along with the presence of new devices (e.g., smartphones, tablets), HCI has expanded its perceptual concept from interaction with a computer to that of any target device, and it has been incorporated into multiple disciplines, such as computer science, cognitive science, and human-factors engineering.

Having a good HCI design is a challenging task since it not only has to cope with ``more than just a computer now... but considerable awareness that touch, speech, and gesture-based interfaces'' \cite{churchill2013teaching} but also requires a substantial cognitive effort to think and make the product that encompassed the aspects of \emph{useful, usable}, and ultimately \emph{used} by the public \cite{idf2020}, such as user interfaces assisting monitoring and system operational tasks~\cite{dang2020agasedviz,le2019visualization}. Literature work has contributed significantly to fostering HCI from imposing design principles, processes, guidelines to teaching. For example, the ACM SIGCHI Executive Committee~\cite{hewett1992acm} developed a set of curriculum recommendations for HCI in education, or Ben Shneiderman~\cite{shneiderman2016designing} suggested eight golden rules of interface design. Adopting the existing guidelines, teachers/instructors have attributed an abundance of efforts to support students in understanding the concept, developing, and creating a good interaction design in classrooms \cite{churchill2013teaching,greenberg1996teaching}. In line with instructing learners, curriculum, teaching/learning methods, project outputs, tools/techniques are reported to share with the HCI community. In this regard, findings are often observed from the instructors' perspectives, the questions of \emph{how}, and \emph{to what extent} students perceive the design principles are unexplored.

Having the answers to these questions would play an essential indicator for both educators and learners as it allows them to reshape their perceptual thinking on how HCI is being taught and learned. For instructors, looking at HCI from students' perspectives enables them to reorganize teaching materials and methods so that learning performance can be best achieved. For learners, the opportunity of having their point of view to justify or being justified by instructors/peers would allow them to develop the critical thinking skills prepared for their future careers.

To the best of our knowledge, no work in the literature exploits these questions in the HCI domain, making this research a unique contribution. In this study, we seek to answer the aforementioned questions by decomposing them into sub-questions as 1) Given a set of design principles/guidelines, to what extents students follow them, 2) Which part of the HCI design the learners focus on, and 3) Do they have the same perspectives on adopting design principles and are these views consistent? By addressing these research questions qualitatively and quantitatively, the contributions of our paper can be laid out as:
\begin{itemize}
    \item it reports the instructional methods and instruments used for teaching and learning HCI in the classroom.
    \item it provides an analysis of the qualitative method for student peer-review project assessment.
    \item it extracts insights of HCI design principles adoption in the classroom.
\end{itemize}
The rest of this paper is organized as follows: Section~\ref{sec:related_work} summarizes existing research that is close to our paper. Section~\ref{sec:method} presents study design methods for collecting students' information. Section~\ref{sec:result} analyzes data and provide insights in detail. Section~\ref{sec:conclusion} concludes our paper with future work direction.
%% Related work section %%%%%%
\section{Related work}
\label{sec:related_work}
Numerous researches on teaching HCI guidelines have been studied: from the general design of the course, major topics should be covered to the incorporation of user-centered design. One of the fundamental literature in designing the HCI curriculum is presented in 1992 by Hewett et al.~\cite{hewett1992acm}, ``ACM SIGCHI Curricula for Human-Computer Interaction''. The report provided ``a blueprint for early HCI courses''~\cite{churchill2013teaching}, which concentrated on the concept of ``HCI-oriented,'' not ``HCI-centered,'' programs. Therefore, it is beneficial to frame the problem of HCI broadly enough to aid learners and practitioners to avoid the classic pitfall of design separated from the context of the problem~\cite{hewett1992acm}.  Regarding the in-class setting, PeerPresents~\cite{PeerPresents} introduced a peer feedback tool using online Google docs on student presentations. With this approach, students receive qualitative feedback by the end of class. To encourage immediacy in the discussion, our system provides feedback on-the-fly: the presenters receive feedback and questions as they are submitted to the system, the presenters can start the discussion right after their presentation session. Instant Online Feedback~\cite{figl2009students} demonstrates the use of instant online feedback on face-to-face presentations. Besides the similar feedback form containing quantitative and qualitative aspects, we include the demo of the final interface within the form so that the audience can directly perform testing. Online feedback demonstrates its advantages in terms of engagement, anonymity, and diversity. Online Feedback System~\cite{hatziapostolou2010enhancing} facilitates students' motivation and engagement in the feedback process. Compared to the face-to-face setting, the establishment of anonymity encourages more students to participate and contributes to balanced participation in the feedback process~\cite{figl2009students}. Along with a higher quantity, diversity in the feedback elicits novel perspectives and reveals unique insights~\cite{ma2015exiting}. For a classroom setting, time is often limited within a class session. With such time restriction, the time frame dedicated to giving feedback should be held rather low~\cite{figl2009students}. Peer feedback should be delivered in a timely manner~\cite{kulkarni2015peerstudio}: for online classes, the feedback only demonstrates its efficiency if provided within 24 hours. Another study shows that instant feedback helps enhance accuracy estimate than that received at the end of a study because of its immediacy~\cite{magrabi2005general}.

%% Methods section %%%%%%
\section{Materials and Methods}
\label{sec:method}

\subsection{Teaching method and activities}
The main goal of our study is to investigate the adoption of HCI design principles in the classroom setting from the learners' perspective. It is a daunting task for instructors to justify the adoption level due to variations in visual cognitive styles. 
For example, teachers may like bright and high contrast design, whereas students prefer a colorful dashboard. 
Thus, a teaching method and assessment should be carefully designed to help motivate and engage students and adequately reflect their performance level. 
% In our study, we consider that students' motivation and engagement are the most critical factors attributed to learning outcomes. 
Literature work has proposed several teaching methods~\cite{carter2011review}, such as inquiry-based learning, situated-based learning, project-based learning, and studio-based learning. 
% Each method has its advantage and limitation depending on the course's objective. 
As such, we chose the studio-based learning approach due to its characteristics, encompassed our critical factors. These characteristics can be laid out as follows:

\begin{itemize}
    \item Students should be engaged in project-based assignments (\textbf{C1}). 
    \item Student learning outcomes should be iteratively assessed in formal and informal fashion through design critiques (\textbf{C2}).
    \item Students are required to engage in critiquing the work of their peers (\textbf{C3}).
    \item Design critiques should revolve around the artifacts typically created by the domains (\textbf{C4}).
\end{itemize}

Based on the above criteria, we divided the entire course work into four main activities spanning over 16-week long, including 1) Introduction to HCI, 2) Homework assignment, 3) Projects engagement: Throughout the course, students were actively involved in two projects (characteristic \textbf{C1}), and 4) Evaluation and design critiques: In each assignment/projects both instructors and students were engaged for evaluations (characteristics \textbf{C2, C4}), this process not only helps to avoid bias in the results but also allows learners to reflect their learning for justifying their peers (characteristic \textbf{C3}). The assignment's purpose is to give students some practice in the design of everyday things. In the first project, we incorporated the problem-based learning method on the problem extended from the assignment. The second project had a different approach by adapting the computational thinking-based method, where students can explore a real-life problem and then solve it on their own.

\subsection{Participants}
The present study was conducted with university students enrolled in the Human-Computer Interaction course in computer science. There was a total of 83 students, of which 62 students are undergraduate, 13 masters, and eight doctoral students. There were 64 males and 19 females.

\subsection{Assessment tool}
Aalberg and Lors~\cite{aalberg2018active} indicated that `the lack of technology support for peer assessment is one likely cause for the lack of systematic use in education.' To address these issues, many assessment tools have been introduced  \cite{figl2009students,gehringer2001electronic,PeerPresents}. However, no such a comprehensive tool enables instructors to carry out a new task that has not been presented. Particularly, in our study, the evaluation requires features such as instant feedback, timing control, list preview, or even 3D objects embedded applications. We developed a standalone tool to facilitate peer evaluation and data collection, supporting multiple application types, including webpage, web-based VR, and AR. The tool is expected to serve these goals: 

\begin{itemize}
    \item \textbf{G1} From the presenter's perspective: Each presenter (group or individual) demonstrates their interface design within their session. The presenter can see the feedback and evaluation visualization within their session.
    
    \item \textbf{G2} From the audience's perspective: The audience can give comments and evaluations for the current design presented~\cite{gehringer2001electronic}. Authentication for the in-class audience is necessary for input validity. 
    
    \item \textbf{G3} The system presents online, instant evaluation from the audience (anonymously) to the presenter without interrupting the presentation.
\end{itemize}{}

\begin{figure}[t]
\includegraphics[width=\textwidth]{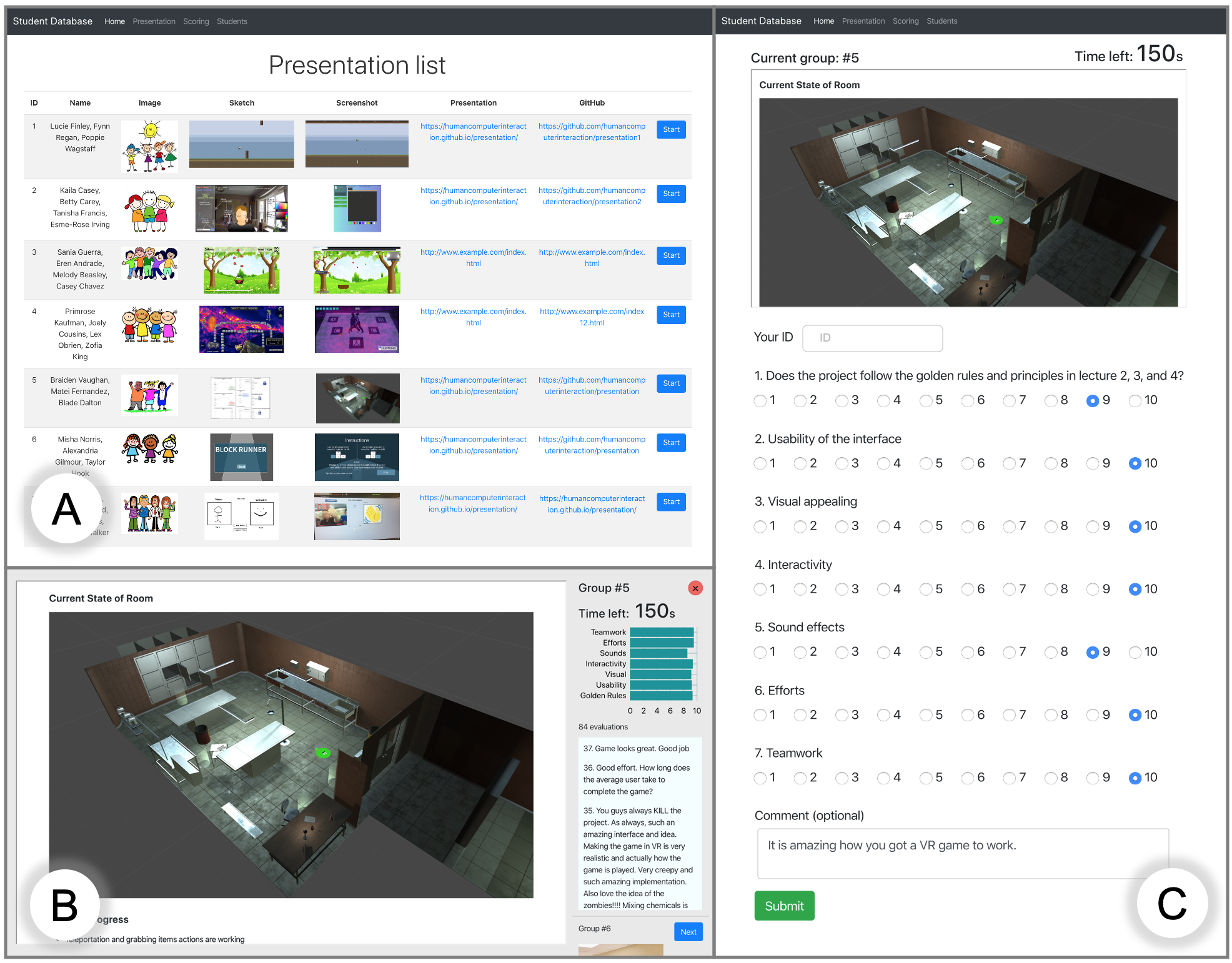}
\caption{User interface of the web presentation application in various contexts. A) List of presentations from the instructor's view, B) Presenter's view, C) Audience's view.}
\label{fig:web}
\end{figure}

\subsubsection{User Interface}
Figure \ref{fig:web} describes our web presentation system according to three perspectives. Panel A shows the presenter list (student names and group members' images have been customized for demonstration purposes). This view allows the instructors to manage students' turn to present, each group has two thumbnail images for sketch and final design, indicating the development process. Panel B is the presenter's view with live comments on the right-hand side, updated on-the-fly. The average scores and evaluation are updated live in the presenter's view -- visualized in an interactive, dynamic chart. Panel C is the audience's view for providing scores and feedback on the presenting project. 

Each presentation is given a bounded time window for demonstration and discussion. The system automatically switches to the next presentation when the current session is over, renewing both Panel B and C interfaces. 
During the presentation, the clock timer in the two panels are synchronized, and the assessment and comments submitted from Panel C are updated live on Panel B. 
Questions for peer assessment consist of seven 10-point Likert scale questions (which ranged from ``strongly disagree(1)'' to ``strongly agree(10)'') and one open-ended question. The criteria for scoring are as follows.

\begin{itemize}
    \item \textbf{Q1}, Does the interface follow the \textbf{Golden rules} and principle in design~\cite{shneiderman2016designing}?
    \item \textbf{Q2}, \textbf{Usability}: The ease of use of the interface~\cite{webvr}.
    \item \textbf{Q3}, \textbf{Visual appealing}: Is the design visually engaging for the users~\cite{sims1997interactivity}?
    \item \textbf{Q4}, \textbf{Interactivity}: To what extent does the interface provide user interactions~\cite{sims1997interactivity}?
    \item \textbf{Q5}, \textbf{Soundness}: The quality and proper use of the audio from the application~\cite{sims1997interactivity}.
    \item \textbf{Q6}, \textbf{Efforts}: Does the group provide enough effort for the work?
    \item \textbf{Q7}, \textbf{Teamwork}: Is the work equally distributed to team members?

\end{itemize}

\subsection{Assignment and Project outputs}
\label{assignmentOutput}

\begin{figure}
\includegraphics[width=\textwidth]{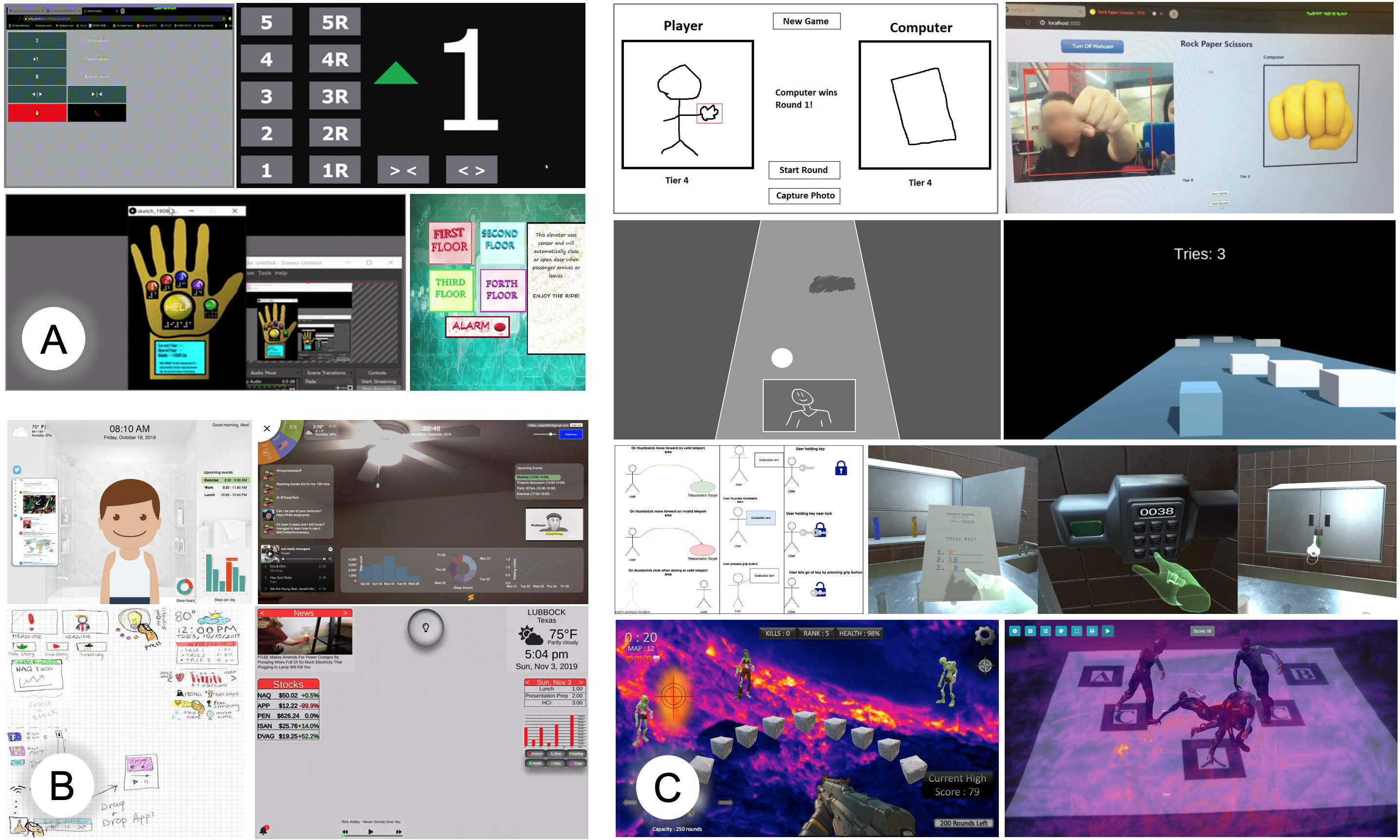}
\caption{Selected work on interface design for the individual assignment (A), project 1 (B), project 2 (C). For each project 1 and 2, sketches are on the left, and final designs are on the right. Our study covers a variety of applications, including Object Detection (as a Computer Vision application), Web-based 3D application, VR, and AR. The interdisciplinary nature is presented in the final result.}
\label{fig:collage}
\end{figure}

There were a total of 126 visual interaction designs as assignment and project outcomes, including 82 sketches from the assignment, 21 problem-based designs, and 22 computational thinking based solutions. Figure \ref{fig:collage} presents some selected work corresponding to the assignment, project 1, project 2, respectively.

\begin{figure}
\includegraphics[width=\textwidth]{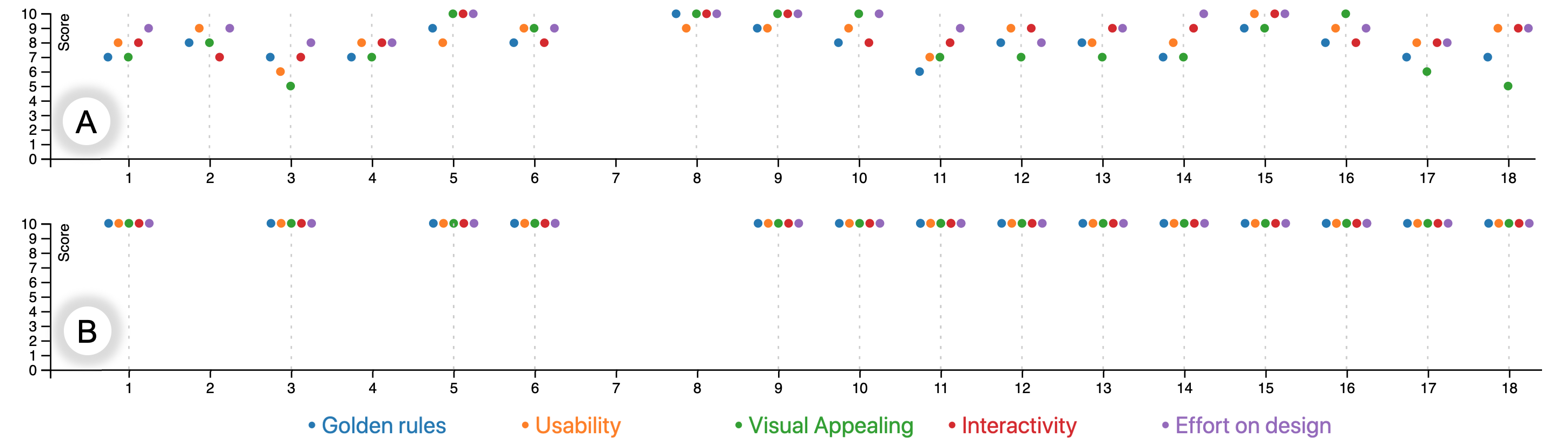}
\caption{Examples of good (panel A) and inadequate (panel B) peer grading for 18 students. Each dot represents grade (from 1 to 10), corresponds to five criteria above.}
\label{fig:grading}
\end{figure}

Previous work showed that there was some ambiguity in the results when students gave scores to their peers. Part of the issue is that they wanted to be `nice' or aimed to get the job done. To alleviate these issues, unusual score patterns will be excluded, and evaluators (students) will be notified anonymously through their alias names. Examples of good grading and inadequate grading are presented in Figure~\ref{fig:grading}. Good grading is illustrated via the diversity in assessment outcome among different criteria, opposite to inadequate examples.

%% Result section %%%%%%
\section{Results}
\label{sec:result}
\subsection{R1: Given a set of design principles/guidelines, to what extents students follow them?}

In order to tackle the first research question, we are looking for an indication among the students that reflects the overall level of guideline compliance. To measure the extent that the learners follow the guidelines, we use the typical statistic measures mean and standard deviation, where each record is a quantitative assessment from a user to a presenter, spreading on the provided criteria.

Table~\ref{tab:extent} presents the mean and standard deviation on the peer assessment score for project 1 and project 2. In both projects, \emph{Efforts} criterion always has the highest mean value, indicating that the students highly appreciate the efforts their peers put in. Regarding project 1, \emph{Visual Design} has the lowest mean (8.20) and also the highest standard deviation (1.48), showing variation visual cognitive styles and diverse individual opinions on what should be considered ``visually engaging''. In project 2, \emph{Golden Rules} criterion has the mean increased and became the criterion with the lowest standard deviation (1.12), meaning that there is less difference between the perception of Golden Rules among the students, demonstrating that the students present better compliance to the guidelines. As an experience learned from project 1, \emph{Sounds} was introduced in project 2 as a channel for feedback interactions. \emph{Sounds} having the lowest mean (7.33) and highest standard deviation (2.30), setting it apart from other standard deviations (around 1.1 and 1.2), indicating that the students had difficulty incorporating this feature on their projects. Overall, the other criteria have their mean values increased from project 1 to project 2, showing a better presentation and understanding of the guidelines.

\begin{table}
\caption{Mean and standard deviation on peer assessment score for project 1 and 2}
\label{tab:extent}
 \scriptsize%
	\centering%
\begin{tabu}{@{}llrr@{}}
\toprule
          & Criteria        & \multicolumn{1}{l}{Mean} & \multicolumn{1}{l}{Standard deviation} \\ \midrule
Project 1 &                 & \multicolumn{1}{l}{}     & \multicolumn{1}{l}{}                   \\ \midrule
          & Golden Rules    & 8.39                    & 1.30                                  \\
          & Efforts         & 8.90                    & 1.23                                  \\
          & Interactivity   & 8.56                    & 1.27                                   \\
          & Usability       & 8.45                    & 1.31                                  \\
          & Visual Design & 8.20                    & 1.48                                  \\ \midrule
Project 2 &                 & \multicolumn{1}{l}{}     & \multicolumn{1}{l}{}                   \\ \midrule
          & Golden Rules    & 8.52                    & 1.12                                  \\
          & Efforts         & 8.78                   & 1.24                                 \\
          & Interactivity   & 8.60                    & 1.23                                  \\
          & Usability       & 8.54                     & 1.17                                  \\
          & Visual Design & 8.30                    & 1.36                                  \\
          & Sounds          & 7.33                    & 2.30                                 \\ \bottomrule
\end{tabu}
\end{table}

\subsection{R2: Which part of the HCI design the learners focus on?}
We answer this question using a qualitative method for analysis; we take inputs from students' comments for each project. We expected that the majority of the captured keywords would be centered around UI design, as indicated in previous literature that most of the HCI class focus on UI design~\cite{greenberg1996teaching,hewett1992acm}.

\begin{figure}
\includegraphics[width=\textwidth]{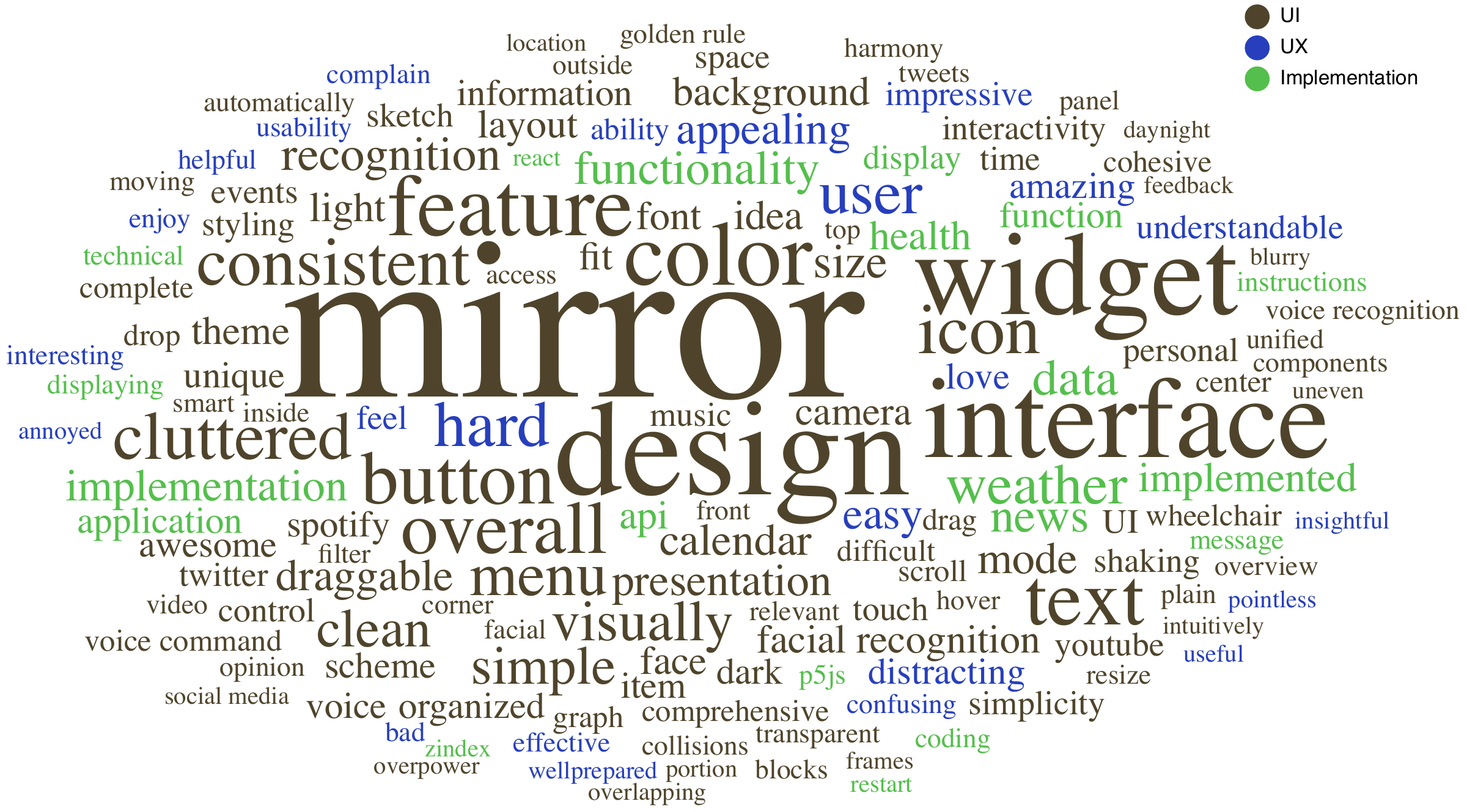}
\caption{Wordcloud constructed from students' comments about the design for project 1. This is the top 150 most common words used, distributed in three categories: UI, UX, and Implementation.}
\label{fig:wordcloud}
\end{figure}

We constructed a wordcloud of the most frequent words in regards to course content, with stop words removed. Figure~\ref{fig:wordcloud} presents the most frequent words used in the project 1 -- interface design for a smart mirror. Bigger font size in the wordcloud indicates more frequent occurrences, hence more common use of the words. Essentially, wordcloud gives an engaging visualization, which can be extended with a time dimension to maximize its use in characterizing subject development~\cite{nguyen2019WordStream}, being able to provide insights within an interactive, comprehensive dashboard~\cite{EQSA}. Hereafter, the number in parentheses following a word indicates the word's frequency. In terms of dominant keywords, besides the design topic such as \emph{mirror}  (72), \emph{design}  (40), \emph{interface}  (33), the students were interested in the service that the interface provides: \emph{widget}  (34) and \emph{feature}  (23), then \emph{color}  (24), \emph{text}  (20) \emph{button}  (19) -- the expectation on fundamental visual components aligned, with \emph{cluttered} (16) and \emph{consistent} (16) -- highlighting the most common pitfall and standard that the design should pay close attention. Functionality in design are taken into account: \emph{touch} (4), \emph{draggable}  (8), \emph{facial recognition}  (6), \emph{voice command}  (3). As such, when looking at an application, students focused primarily on UI design, hence our study confirmed existing research~\cite{greenberg1996teaching,hewett1992acm}. 

Furthermore, user experience (UX) concerns are demonstrated through the students' perspective. By exploring more uncommon terms, user's experience is indicated by: \emph{understandable}  (4), \emph{usability}  (2), \emph{helpful}  (2). Users' emotions and attitudes towards the product are expressed: \emph{easy}  (8), \emph{love}  (5) and \emph{enjoy}  (2) (positive), in contrast to \emph{hard}  (15), \emph{distracting}  (5), \emph{difficult}  (3), \emph{confusing}  (2)  (negative). Indeed, the comments reflect the views of students: \emph{``The mirror is well done. The entire thing looks very consistent and I enjoy the speaking commands that lets you know what you are doing.''} (compliment), \emph{``Accessing the bottom menu to access dark mode could be a little confusing for some users as no icons are listed on the screen.''} (suggestion).
% The feedback was straightforward and concentrated on the overall graphic layout and interactivity of the design. 
The findings on UX can complement existing work in ways that emphasize the need to integrate UX subjects in the HCI curriculum. HCI principles can be considered a crucial instrument for UX development; these results validated that HCI is the forerunner to UX design~\cite{idf2020}.

Overall, we can classify the keywords into three groups: UI, UX and Implementation. Besides UI and UX discussed above, the Implementation aspect is viewed through \emph{data} (10),  \emph{implementation} (8), \emph{api} (6), \emph{function} (5), and
\emph{coding} (2). 
Compared to the expectation, there is a minimal amount of terminologies to the principle used, such as \emph{golden rule} (2) -- with only two occurrences, as opposed to the large number of visual elements mentioned in practical development. Some students even suggested \emph{wheelchair} (3), as they consider the design for a variety of users. This pattern demonstrates that the focus shifts from theoretical design to direct visual aesthetics, as the students perceived and adopted the HCI principles effectively to apply them in the empirical application. To sum up,
the keywords retrieved from the wordcloud cover primarily UI and UX interests from the students' perspective in the process of adopting HCI into interface design, demonstrating the diversity and broad coverage that peer assessment outcomes can provide. The method can be scaled to other disciplines that regard crowd wisdom in the development process.

\subsection{R3: Do they have the same perspectives on adopting the design principles and are these views consistent?}

To answer the third question, we are looking for an agreement among students when they evaluated their peers. We hypothesize that the HCI principles are adopted when the scores provided by learners are \emph{agreed}, and this agreement may imply that students have the same level of understanding of the principles.

To measure the level of agreement among learners, we use the intraclass correlation coefficient (ICC) \cite{mcgraw1996forming}, where each student is considered a rater, the project is the subject of being measured. ICC is widely used in the literature because it is easy to understand, can be used to assess both relative and absolute agreement, and the ability to accommodate a broad array of research scenarios compared to other measurements such as Cohen's Kappa or Fleiss Kappa \cite{nichols2010putting}. Cicchetti \cite{cicchetti1994guidelines} provides a guideline for inter-rater agreement measures, which can be briefly described as poor (less than 0.40), fair (between 0.40 and 0.59), good (between 0.60 and 0.74), excellent (between 0.75 and 1.00).

\begin{table}[!ht]
  \caption{Inter-rater agreement measures for project 1 and project 2.}
  \label{tab:agreement}
  \scriptsize%
	\centering%
  \begin{tabu}{%
	{c}%
	{c}%
	{c}%
	{l}
	{c}
	{c}
	{r}
	}
     \toprule
    & Subjects & Raters &  Type & Agreement & Consistency & ICC scale \cite{cicchetti1994guidelines}  \\  
   \midrule
    Project 1 & &&&&&\\
      \midrule
	 & 19 & 77 &  Golden Rules & 0.963 & 0.972 & excellent  \\  
     & 19 & 77 &  Efforts & 0.900 & 0.970 & excellent  \\  
    & 19 & 77 &  Interactivity & 0.955 & 0.962 & excellent  \\  
    & 19 & 77 &  Usability & 0.965 & 0.972 & excellent  \\  
    & 19 & 77 &  Visual Design & 0.971 & 0.977 & excellent  \\  
     Project 2 & &&&&\\
      \midrule
	& 21 & 75 &  Golden Rules & 0.922 & 0.947 & excellent  \\  
     & 21 & 75 &  Efforts & 0.932 & 0.952 & excellent  \\  
    & 21 & 75 &  Interactivity & 0.896 & 0.923 & excellent  \\  
    & 21 & 75 &  Usability & 0.879 & 0.886 & excellent  \\  
    & 21 & 75 &  Visual Design & 0.964 & 0.971 & excellent  \\  
    & 21 & 75 &  Sounds & 0.970 & 0.977 & excellent  \\ 
    \bottomrule
  \end{tabu}%
\end{table}

Table~\ref{tab:agreement} provides the results on the level of agreement and consistency when students evaluate their peers. There is a difference in the number of raters (83 students in total, 77 in project 1, and 75 in project 2); this is due to the exclusion of ambiguous responses, as noted in Section~\ref{assignmentOutput} and Figure~\ref{fig:grading}. It can be seen from Table~\ref{tab:agreement} that students tend to give the same score (agreement score = 0.963) given the principle guideline (or golden rules) in project 1, so do as in project 2 (agreement score = 0.922). These scores are considerably high (excellent) when mapped to the inter-rater agreement measure scales suggested by Cicchetti~\cite{cicchetti1994guidelines}. In addition, we also find a high consistency among students when evaluating the other aspects of the projects such as efforts, interactivity, usability, and visual design since consistency scores are all above $0.75$.

%% Conclusion section %%%%%%
\section{Conclusion and Future work}
\label{sec:conclusion}
In this paper, we have presented the learners' perspective on the perception and adoption of HCI principles in a classroom setting. A standalone web presentation platform is utilized to gather instant online peer feedback from students throughout the course. The qualitative analysis of peer feedback demonstrated that the students primarily emphasize the design feature and then visual components and interactivity.
This approach can be extended to present subject evolution and student development over time, providing a bigger context. 
The outcome also pointed out that the interests spread in both UI and UX, suggesting further incorporation of UX in the HCI curriculum, which is currently having a shortage in the existing literature. On the quantitative analysis, we have found that the students followed the guidelines by a large margin. The results indicated a high level of agreement among the students, determined by the inter-rater agreement measures. We also have found high consistency within the class regarding other aspects evaluated, such as efforts, interactivity, usability, and visually appealing. 

This research, however, is subject to several limitations. The course duration in this study is short, with only 16 weeks; a more prolonged period can help reduce random errors. Another limitation is the sample size, with 83 students; we believe that a larger sample size could increase the generalizability of the result (nevertheless, our sample size meets the minimum requirement suggested by~\cite{albano2017introduction}). We will expand the study in the following semesters for future work, and we expect that researchers can conduct related studies to confirm our findings.

% % ---- Bibliography ----
%

\bibliographystyle{splncs04}
\bibliography{reference}

\end{document}